\documentclass[useAMS,usenatbib]{mn2e}
\usepackage{graphicx}
\usepackage{amssymb}
\usepackage{epstopdf}
\usepackage{amsmath}

\usepackage[usenames,dvipsnames]{color}

\def\etal{{et al.~}}

\def\kms{\>{\rm km}\,{\rm s}^{-1}}

\def\Msun{\>{\rm M_{\odot}}}

\title[SMBH growth parameters]
  {SMBH growth parameters in the early Universe of Millennium and Millennium-II simulations}
\author[Majda Smole, Miroslav Micic, Nemanja Martinovic]
  {Majda Smole,$^1$ \thanks{E-mail:msmole@aob.rs,micic@aob.rs,nmartinovic@aob.rs}
   Miroslav Micic,$^1$ Nemanja Martinovi\'c,$^1$  \\ 
$^1$ Astronomical Observatory, Volgina 7, 11060 Belgrade 38, Serbia}
  
\date{Released xxxx Xxxxx XX}

\pagerange{\pageref{firstpage}--\pageref{lastpage}} \pubyear{2002}

\def\LaTeX{L\kern-.36em\raise.3ex\hbox{a}\kern-.15em
    T\kern-.1667em\lower.7ex\hbox{E}\kern-.125emX}

\begin{document}

\label{firstpage}

\maketitle

\begin{abstract} We make black hole (BH) merger trees from Millennium and Millennium-II simulations
to find under what conditions $10^{9}\Msun$ SMBH can form by redshift $z=7$. In order to exploit 
both: large box size in the Millennium simulation; and large mass resolution in the Millennium-II
simulation, we develop a method to combine these two simulations together, and use the Millennium-II 
merger trees to predict the BH seeds to be used in the Millennium merger trees. We run multiple 
semi-analytical simulations where SMBHs grow through mergers and episodes of gas accretion 
triggered by major mergers. As a constraint, we use observed BH mass function at 
redshift z=6. 
We find that in the light of the recent observations of moderate 
super-Eddington accretion, low mass seeds ($100\Msun$) could be the progenitors 
of high-redshift SMBHs ($z\sim7$), as long as the accretion during the accretion
episodes is moderately super-Eddington, where ($f_{\mathrm{Edd}}=3.7$)
is the effective Eddington ratio averaged over 50 Myr.

\end{abstract}


\section{INTRODUCTION}

Super-massive black holes (SMBHs) with masses $10^{6}\Msun$ to $10^{10}\Msun$ populate 
centres of spiral galaxies with central bulge and centres of massive elliptical galaxies
\citep{korm}. These SMBHs grow through mergers and episodes 
of gas accretion.  During accretion phase, black holes (BHs) are indirectly observed 
as active galactic nuclei (AGNs) or quasars, where quasars powered by SMBHs with masses $\sim 10^{9}\Msun$ 
were observed as early as $z\sim6$ (e.g., \citealt{will}; \citealt{fan06}; 
\citealt{rosa}). Those observations showed that BHs had very little 
time ($<1\rm{Gyr}$) to grow from initial seeds \citep{fan01}. An extreme example is SMBH with 
mass $2\times10^{9}\Msun$ detected at $z=7.085$ \citep{mort}. 
It would take a $10^{5}\Msun$ BH seed and continuous 
(many e-folding times) gas accretion at the Eddington limit, to grow such 
massive BH so early. Starting with lower mass BH seeds would 
require super-Eddington accretion. Obvious question is: how massive were the BH seeds from which 
SMBHs formed so early and what are the key SMBH growth parameters?   
%\subsection{Pop III stars}

Several mechanisms for BH seeds formation have been proposed. We will 
briefly discuss the three most popular mechanisms.

Pop III stars, the first 
metal-free stars that started to form at $z\sim20$, might have left BH 
seeds that grew to SMBHs observed in the first 
billion years after the Big Bang (\citealt{mr}; \citealt{heger}; 
\citealt{wise}).
Masses of those BH seeds depend on the masses of Pop III stars,
which could be in range 
$60-300\Msun$ \citep{volker}, or even 
$1000\Msun$ \citep{hirano}.
Some simulations showed 
that Pop III stars were not so massive and that they formed in binary 
systems or in clusters (\citealt{turk}; \citealt{clark}; \citealt{stacy}), 
but other authors showed that Pop III stars were formed isolated 
(\citealt{on}; \citealt{hirano}). It is typically assumed in the 
literature that Pop III stars form BH seeds with masses close to $100\Msun$.

Previous attempts to form $\sim$ 10$^9\Msun$ SMBHs at $z\sim6$ from 
$100\Msun$ seeds required continuous accretion close 
to or exceeding the Eddington limit and radiative efficiencies of
$\epsilon\lesssim0.1$ (\citealt{hl}; \citealt{tyler}; \citealt{vb}; 
\citealt{daniel};  \citealt{john1}). Even with combination of BHs mergers and accretion,  
low-mass BH seeds must grow at the Eddington limit 
for a significant fraction of time or they need to have early stages of 
super-Eddington accretion in order to explain the observed high-redshift 
quasars (\citealt{yoo}; \citealt{vr}; \citealt{li}; \citealt{pelu}; 
\citealt{sijacki}; \citealt{tanaka1}; \citealt{tanaka2}; \citealt{madau}). 
There are theoretical uncertainties whether BH seeds can sustain such high 
accretion rates for a long time (\citealt{milosavljevic}, \citealt{alva}, 
\citealt{jeon}). It is even more difficult to form $\sim$ 10$^9\Msun$ 
SMBHs at $z\sim7$ considering that the time available for accretion is even
smaller than at $z\sim6$.

%\subsection{Direct collapse}

Another possible mechanism for BH seed formation is direct collapse of 
gas into BH. This process produces BH seeds with masses in range 
$\sim 10^{4}-10^{6}\Msun$ (\citealt{loeb}; \citealt{el}; 
\citealt{oh}; \citealt{bromm}; \citealt{kou}; \citealt{begel1}, \citealt{lodato}, 
\citealt{begel2}). It has been shown that massive BH seeds formed by direct collapse can reach 
$10^{9}\Msun$ at $z>6$ (e.g., \citealt{dij}; \citealt{aga}; 
\citealt{bonoli}; \citealt{petri}; \citealt{john1}). However, BH seeds in 
these models may still need to accrete at nearly the Eddington limit for 
a significant fraction of time. The main challenge in direct collapse models 
is to avoid fragmentation and star formation. Fragmentation will not occur 
if gas has no mechanism for efficient cooling, that is if there are no metals and if formation of $\mathrm{H}_{2}$ molecules is 
prevented via strong UV background. Only extremely rare haloes satisfy this 
condition. Direct collapse is expected to occur in haloes with virial 
temperature $T_{\mathrm{vir}}\gtrsim10^{4} \mathrm{K}$ and masses 
$10^{7}-10^{9}\Msun$ at $10<z<20$. Recently, models have 
been proposed where direct collapse might be possible even without a 
strong UV radiation background (\citealt{ina}; \citealt{tl}), but \cite{eli} 
have shown that those models are not viable because molecular cooling will still 
occur, as the gas density increases, which leads to fragmentation.
Alternatively, major mergers may lead to a rapid gas inflow. In such case, 
turbulence may be the inhibitor of fragmentation, and the requirement of
metal-free gas may be relaxed \citep{mayer}.

The third possible mechanism for BH seed formation is collapsing star 
clusters. \cite{deve} suggested model where mergers of Pop II stars 
lead to the formation of a very massive star which collapses into a BH 
with mass $\sim10^{3}\Msun$, at $z\sim10-20$. \cite{davi} 
proposed a model of a star cluster which contains only BHs and main 
sequence low mass stars. More massive BHs sink to the centre and form BH 
cluster which further collapses and forms massive BH with mass 
$10^{5}\Msun$, at $z>10$. These seeds can grow via 
mergers and accretion to form SMBHs. Some authors showed that such seeds 
could explain observed quasars at $z<5$ but have trouble explaining quasars 
at higher redshifts (\citealt{ebi}; \citealt{vh}; \citealt{islam}).

BH mergers could contribute to the SMBH growth. \cite{john2}
argued that growth of the most massive BH at high redshift is solely due to 
the accretion and that mergers can be neglected. Their conclusion is based 
on: the fact that only a small number of high-redshift haloes are capable of hosting Pop III stars; 
Pop III star formation rate is reduced due to Lyman-Werner (LW) radiation 
field; and the results of the recent large-scale cosmological simulations tracking the buildup 
of SMBH, where it is claimed that mergers do not affect the growth of SMBHs 
\citep{degraf}. Simulation by \cite{degraf} does show that negligible fraction
of high-redshift SMBH mass is acquired in mergers but this is a consequence
of the poor mass resolution. Their simulation does not form haloes smaller than $5\times
10^{10}h^{-1}\Msun$. While it is true that Pop III star formation can not 
take place in smallest haloes, it does occur in haloes with $T_{\rm vir}\sim1000-2000\rm{K}$ 
 which corresponds to $M_{\rm halo}\gtrsim10^8\Msun$ at 
redshift $z>6$. Also, haloes with $M_{\rm halo} \gtrsim10^8 \Msun$ are most
likely self-shielded from LW radiation due to the large $\rm{H}_2$ column density \citep{wise2008}.
Hence, it is clear that haloes with mass 
$10^8 \Msun <M_{\rm halo} \lesssim10^{10} \Msun$ host Pop III stars
and if these stars produce massive BHs, then their mergers should
be relevant to the growth of high redshift SMBHs.

We investigate if light BH seeds ($100\Msun$) planted into 
haloes of Millennium simulation ($M_{\rm halo}>~10^{10}\Msun$) 
and Millennium-II simulation ($M_{\rm halo}>~10^8\Msun$) can grow
into SMBHs that have been observed at $z\sim7$.

We also investigate if gravitational wave recoil could prevent the formation of SMBH. 
During a BH merger gravitational wave radiation is produced. Asymmetric 
emission of gravitational radiation can lead to BH kick. Gravitational 
waves carry a non-zero net linear momentum, which establishes a preferential 
direction for the propagation of the waves and the centre of mass of the 
binary recoils in the opposite direction \citep{redmount}. The magnitude 
of the gravitational wave recoil depends on the mass ratio of BHs, 
the spin magnitude and orientation with respect to the binary orbital 
plane and the eccentricity of the orbit (\citealt{campa}; \citealt{schni}; 
\citealt{baker}). Gravitational wave recoil can eject a newly formed BH 
from the host halo if the BH speed is larger than the escape velocity from 
the halo centre. BHs can be kicked with a speed as large as  
$\sim4000 ~\mathrm{km~s}^{-1}$ in special orbital configurations 
(\citealt{herrman}; \citealt{gonzalez}; Campanelli \etal 2007; 
\citealt{schni}; \citealt{kopp}). At high redshift, dark matter haloes generally 
have small masses and thus small escape velocities, so BH with speed 
$\geq150 ~\mathrm{km~s}^{-1} $ can be ejected even from the most massive 
haloes at redshift $z\geq11$ (\citealt{merritt}; \citealt{micic2006}; 
\citealt{volonteri2007}; \citealt{schni2007}; \citealt{sesana}; 
\citealt{volonteri2010}; \citealt{micic2011}). This effect may play a 
major role in suppressing the growth of SMBH through mergers.

% parametri akrecije

\subsection{Growth Parameters and Their Values}
Large scale structure formation and galaxy dynamics lead to BH mergers. If those 
mergers are ignored, BH growth depends on three gas accretion parameters. 
Those parameters are radiative efficiency, Eddington ratio and the time 
that a BH spends accreting. 

%\subsection{Eddington ratio}

Radiative luminosity of a BH, $L$, which is accreting at a rate 
$\dot{M}_{\mathrm{BH}}$,  is given by ${L=\epsilon \, \dot{M}_{\mathrm{BH}} \,c^{2}/(1-\epsilon)}$, 
where $c$ is the speed of light and $\epsilon$ is the radiative efficiency. 
Eddington ratio is $f_{\mathrm{Edd}}=\frac{L}{L_{\mathrm{Edd}}}$, where 
${L_{\mathrm{Edd}}=1.26\times10^{38}(\frac{M_{\mathrm{BH}}}{\mathrm{M}_{\odot}}})[\mathrm{erg}~\mathrm{s^{-1}}]$ 
is Eddington luminosity. The accretion rate at which a BH will radiate at a given Eddington ratio is given by:

\begin{equation}
\noindent \dot{M}_{\mathrm{BH}}=\dfrac{(1-\epsilon) \, f_{\mathrm{Edd}}\,L_{\mathrm{Edd}}}{\epsilon c^{2}}.
\end{equation}

\noindent After integration the final BH mass $M_{\mathrm{BH}}$, as a function of its initial mass $M_{\mathrm{BH,0}}$ is:

\begin{equation}
M_{\mathrm{BH}}=M_{\mathrm{BH,0}}\times \mathrm{exp} \, [\,\dfrac{f_{\mathrm{Edd}}\,(1-\epsilon)}{\epsilon}\,\,\dfrac{t_{\mathrm{f}}-t_{\mathrm{i}}}{t_{\mathrm{Edd}}}]
\end{equation}

\noindent where $t_{\mathrm{Edd}} = 450~\mathrm{Myr}$, $t_{\mathrm{f}}$ i $t_{\mathrm{i}}$ are the ages of the universe when the BH attains its final mass and at the time of seed formation, respectively \citep{john2}.

%\subsection{Radiative efficiency}

Radiative efficiency is the efficiency of conversion of rest-mass into 
energy during accretion and it depends on the BH spin. Radiative efficiency 
can take values from 0.057 for accretion on to Schwarzschild BHs 
to 0.42 for fast rotating Kerr BHs \citep{shapiro}.
Mean value of radiative efficiency 
for quasars can be estimated by comparing the local 
SMBH mass density with the total AGN luminosity 
per unit volume in the Universe integrated over time \citep{soltan}. This is Soltan's argument. Previous 
works based on Soltan's argument showed that the mean value of radiative 
efficiency is $\epsilon\geq0.1$ (\citealt{elvis}; \citealt{yu}; \citealt{davis}).
Some authors have found that radiative efficiency changes with 
the redshift (\citealt{wang}; \citealt{lietal2012}) and increases 
with the mass of the accreting BH (\citealt{davis}; \citealt{shankar}; \citealt{lietal2012}).

%\subsection{Eddington ratio}

It is usually assumed that BH luminosity during accretion cannot 
be greater than the Eddington luminosity. However, evidence for 
super-Eddington accretion has been growing recently.  
\cite{shen} used a sample of $\sim$ 58,000 quasars at $z\sim0.3-5$ from Sloan Digital Sky Survey (SDSS) DR7 
catalogue \citep{dr7} to estimate their Eddington ratios. They found that 
the highest observed Eddington ratio for quasars is $f_{\mathrm{Edd}}\sim3$, 
but that those quasars which radiate above the Eddington limit are rare.
\cite{du} observed three quasars, Mrk 335, Mrk 142 and IRAS F12397+3333, 
using the 2.4-m Shangri-La telescope at the Yunnan Observatory in China.
One of their goals was to measure BH masses and Eddington ratios. 
They found that the lower limits on the Eddington ratios for these objects 
are 0.6, 2.3, and 4.6. \cite{page} have shown that soft X-ray spectrum of 
the highest redshift quasar yet found, ULAS J112001.48+064124.3 at $z=7.085$
\citep{mort}, obtained with $Chandra$ and $XMM-Newton$, suggests that the quasar 
is accreting above the Eddington limit, $f_{\mathrm{Edd}}=~5^{+15}_{-4}$. 
Their findings of moderate super-Eddington accretion are consistent with the 
Eddington ratio estimated from the UV luminosity of that object, 
$f_{\mathrm{Edd}}=1.2^{+0.6}_{-0.5}$ \citep{mort}. In order to fit the 
observed statistics of far-infrared and X-ray spectra of AGNs	
at $z\gtrsim2$ \cite{lapi1} assumed model where Eddington ratio 
depends on the redshift. During the exponential growth of the BH, the 
maximum Eddington ratio is $f_{\mathrm{Edd}}\sim4$ for $z=6$, and 
$f_{\mathrm{Edd}}\sim1$ for $z=2$, with constant radiative efficiency 
$\epsilon=0.15$. 
Similar results have been found by \cite{li2012}, 
who has also explored the possibility of having a short super-Eddington 
accretion followed by a sub-Eddington accretion in order to explain the 
presence of BHs with masses $\sim10^9\Msun$ SMBHs at $z\sim6$. 

A recent analysis of BH scaling relations showed that
the normalization of the BH mass-bulge relation 
should be increased by a factor of 5, from the previously accepted value 
of $M_{\rm{BH}}=0.1\%M_{\rm{bulge}}$ to $M_{\rm{BH}}=0.5\%M_{\rm{bulge}}$. 
This increases the local mass density in BHs by the same factor and 
decreases the required mean radiative efficiency to values that cannot be
reasonably explained in terms of luminous thin-disc accretion. This may be 
evidence for the radiatively inefficient super-Eddington accretion (\citealt{novak}; \citealt{kh}). 
These works have showed that quasars, at least at some point in their evolution, 
can accrete at the super-Eddington luminosities. 
Recent numerical simulations (\citealt{mc}; \citealt{sad}) of super-Eddington 
accretion suggest that those sources are most likely to be characterized by 
strongly collimated outflows or jets. Accretion in these simulations is mildly 
super-Eddington, $f_{\mathrm{Edd}}=1-10$.

Theoretical works by \cite{vr} assumed early stages of super-Eddington 
quasi-spherical accretion estimated using the Bondi-Hoyle formula 
\citep{bondi}. In this case, when the inflow rate is super-critical, the 
radiative efficiency drops. Hence, the Eddington luminosity is not greatly exceeded 
($L_{\mathrm{Edd}}=\epsilon\dot{M}_{\mathrm{Edd}}c^{2}$).
The accretion rate is initially super-critical by a factor of 10 and
then grows up to a factor of about $10^{4}$. They showed that
BH seeds with initial mass of $1000\Msun$ could explain SMBH at
$z\sim6$ if they go through early phases of super-Eddington accretion
and BH mergers.
Similar approach has been recently argued by \cite{vs}. 
They showed that short-lived intermittent episodes of super-Eddington accretion ($f_{\mathrm{Edd}}>10$)
may increase the BH mass by several orders of magnitude in $\sim10^{7}$ years.
The authors assumed slim disc solution where the luminosity during accretion depends
logarithmically on the accretion rate. 
Since ($f_{\mathrm{Edd}}=\epsilon \frac{\dot{M}}{\dot{M}_{\mathrm{Edd}}}$), 
if the effective radiative efficiency is low, accretion rate can be highly super-Eddington while the emergent
luminosity is only mildly super-Eddington and feedback is limited.
\cite{madau} extended these works and showed that light BH seeds ($100\Msun$)
could explain quasars with $10^{9}\Msun$ SMBH observed at $z\geq6$ if they 
have a few episodes of super-Eddington accretion via slim accretion disc. 
Eddington ratio in these works is $\sim1-10$.

\cite{deh} proposed different mechanism for super-Eddington accretion. 
Momentum driven feedback from an accreting BH gives significant orbital energy 
but little angular momentum to the surrounding gas. Once central accretion 
drops, the feedback weakens and gas falls back towards the BH, forming a small scale 
accretion disc. The feeding rates into the disc typically exceed Eddington by factor of a few.

Based on the mentioned works it can be concluded that super-Eddington accretion is inevitable for the growth of SMBHs.

%\subsection{Quasars lifetimes}

Another important question is for how long accretion can be sustained?
Quasars lifetimes can be estimated using Soltan's argument. Amount of matter 
accreted on to quasars during their lifetime, represented by the luminosity 
density due to accretion, should be less than or equal to the space density
of remnant BHs in the local Universe \citep{soltan}. This method 
depends on the value of radiative efficiency. Another way to estimate this 
parameter is to ask what value of quasars lifetime is required if all bright 
galaxies go through a quasar phase. This approach is not sensitive to the
value of radiative efficiency, but it is affected by the assumed value of
Eddington ratio and galaxy mergers \citep{martini}. Computed values for the duration of accretion using this approach are from 
$10^{6}$ to $10^{8} ~\mathrm{yr}$. 

Typical accretion time for majority of
quasars is Salpeter's time \citep{salpeter}. Salpeter's time is e-folding time-scale for SMBH growth and
its typical value is $\sim~50~\rm{Myr}$:

\begin{equation}
t_{\mathrm{s}}=M/\dot{M}=4.5\times10^{7}\,(\dfrac{\epsilon}{0.1})\,(\dfrac{L}{L_{\mathrm{Edd}}})^{-1}. 
\end{equation}

\noindent A single massive BH seed ($10^{5}-10^{6}~\mathrm{M}_{\odot}$) 
needs to continuously accrete at Eddington luminosity for 380-500 Myr (which corresponds 
to 9 - 11 e-folding times) to be able to grow to $>2\times10^{9}\Msun$ at 
redshift $z\geq6$. Light BH seed ($100~\mathrm{M}_{\odot}$) 
requires 19 e-folding times or 840 Myr of accretion. Alternatively, it would 
take 280 Myr at Eddington ratio 3 (also 19 e-folding times). This means that it would be very hard to 
grow SMBH from light seeds at the Eddington limit by $z=6$ and impossible by $z=7$. Some authors assumed such 
long quasar lifetimes in order to match observed quasars number density at high redshifts (\citealt{hl}; 
Tyler \etal 2003; Sijacki \etal 2009; \citealt{tanaka1}; Tanaka \etal 2012; 
\citealt{john1}).

Continuous accretion at or above the Eddington limit might be problematic due to the feedback. 
Radiation and kinetic power in matter outflows could be so strong that once the accreting BH 
gets too big, it blows out all of the gas in the centre of its host galaxy, 
shutting down the accretion on to the BH \citep{coppi}. If continuous growth at 
the Eddington limit cannot be maintained, the main alternative to these models
are short periods of super-Eddington accretion (\citealt{vr}; \citealt{deh}; \citealt{vs}; \citealt{madau}). 
However, observations show that, in most cases, quasars lifetimes are comparable 
with Salpeter's time and Eddington ratios are between 0.1 and 1. Time that 
SMBH spends accreting depends on radiative efficiency, Eddington ratio 
and BH mass \citep{martini}. \cite{yu} showed that quasar lifetime is a 
function of BH mass. They found that the mean lifetime is 
$3-13\times10^{7}~\mathrm{yr}$ for $\epsilon=0.1-0.3$ and $10^{8}<M_{\mathrm{BH}}<10^{9}~\mathrm{M}_{\odot}$.
It is obvious that if one assumes these typical values of the growth 
parameters, it would be impossible to grow SMBH at high redshift even
if we start with seeds as massive as 10$^7 \Msun$. Accretion would have
to be either super-Eddington or prolonged for more than Salpeter's time 
(many e-folding times instead of one).

Our goal is to examine if BH mergers combined with 
accretion episodes in merger trees of Millennium simulation 
\citep{MS} and Millennium-II simulation \citep{boy}
contribute to the growth of $10^9\Msun$ SMBH at $z\sim7$.
If they do, then values of Eddington ratios and numbers 
of e-folding times during the merger induced accretion episodes in merger 
tree branches can be lowered to a more reasonable values. 
Growth of a SMBH in halo that undergoes 
several major mergers does not require continuous accretion for
many e-folding times. 
Since the rate of major mergers increases with redshift, SMBHs in high-redshift quasars
can grow nearly continuously in a sequence of major mergers (Li \etal 2012, \citealt{tanaka2014}).
Every major merger will reset previous accretion episode
and trigger a new one which effectively reduces number of e-folding times.
That in turn could 
increase the likelihood of low-mass BHs as SMBH seeds.

In Section 2 we describe the method. In Section 3 we present our results. 
We summarize and discuss our results in Section 4.

\section{METHOD}

Our goal is to produce SMBH with mass $>10^{9}\Msun$ 
at $z\sim7$ from light BH seeds using publicly
available data from both 
Millennium simulation \citep{MS} and  Millennium-II simulation \citep{boy}.
As a constraint we use comparison between the observed and calculated BH mass function from our model at $z \sim 6$.

\subsection{Millennium simulation and Millennium-II simulation}

Millennium simulation \citep{MS} is a large $N$-body simulation which 
follows $2163^{3}$ particles within a periodic simulation cube of side 
length $L=500h^{-1}~\mathrm{Mpc}$. 
Each simulation particle has mass $8.61\times10^{8} ~\mathrm{M}_{\odot}$.
The $\Lambda\mathrm{CDM}$ cosmology used for the Millennium simulation is:

\vspace{12pt}

\noindent $\Omega_{\mathrm{tot}}=1.0$, $\Omega_{\mathrm{m}}=0.25$, $\Omega_{\mathrm{b}}=0.045$, $\Omega_{\Lambda}=0.75$, $h=0.73$, $\sigma_{8}=0.9$, $n_{s}=1$,

\vspace{12pt}

\noindent where $h$ is the Hubble constant at $z=0$ in units of 
$100 ~\mathrm{km~s}^{-1}~\mathrm{Mpc}^{-1}$, $\sigma_{8}$ is the 
rms amplitude of linear mass fluctuations in 8$h^{-1}~\mathrm{Mpc}$ 
spheres at $z=0$, and $n_{\mathrm{s}}$ is the spectral index of the 
primordial power spectrum.

Millennium-II simulation \citep{boy} uses the same cosmology
and the same number of particles as the 
Millennium simulation, but in a 
five times smaller box ($L=100h^{-1}~\mathrm{Mpc}$) and thus with 125 times better mass resolution.
Each simulation particle in Millennium-II simulation has 
mass $6.885\times10^{6} ~\mathrm{M}_{\odot}$.
Millennium-II simulation uses GADGET-3 code, 
updated version of GADGET code (\citealt{springel1}, \citealt{springel2}).

Current dominant cosmological paradigm is
ΛCDM, which predicts bottom-up mode
of structure formation, that is, small dark matter haloes forming 
first and then merging into larger haloes later in the life of the Universe.

Since Millennium-II simulation has 125 times better mass resolution
than Millennium simulation, it allows as to track BH growth by
mergers of highest redshift low mass haloes which are too small to be resolved 
due to the lower mass resolution of Millennium simulation.
On the other hand, it is questionable if it is possible to produce a $10^9\Msun$ SMBH 
at $z\sim7$ in a simulation box with sides of length that Millennium-II simulation has due to low population of 
highest mass haloes. In order to have both small haloes and large box, we develop 
a method to use BHs produced in merger trees of Millennium-II
simulation, as BH seeds in merger trees of Millennium simulation, which enable us to cover both 
haloes that are forming very early and abundant highest mass haloes later.

\subsection{Combining Millennium-II and Millennium merger trees}

With $100\Msun$ BH seeds placed in Millennium-II haloes we produce BH growth history. From it, because of the higher mass 
resolution of Millennium-II simulation, we have information about earliest halo formation history.

First, we make merger trees 
which track dark matter halo merger history in Millennium-II 
simulation from redshift $z=23.79$ 
to $z=6.2$. 
Halo is defined as self-bound structures with at least 20 particles.
Merger history of BHs corresponds to the merger history 
of haloes assuming that each halo hosts one BH and that two BHs merge 
right after their host halo merge. We have distinguished between minor 
and major mergers. Merger is major if 
$\frac{M_{\mathrm{halo,}1}}{M_{\mathrm{halo,}2}}\geq0.3$ for 
${M_{{\mathrm{halo,}}1}}<{M_{{\mathrm{halo,}}2}}$.
In the case of a 
minor merger, mass of the newly formed BH is a simple sum of the 
previous BH masses. In the case of a major merger, newly formed BH is 
also accreting gas according to the accretion recipe described in 
following subsections. 
In each halo of Millennium-II simulation we place one BH seed with
$100\Msun$ and assume fixed values for radiative efficiency
and accretion time-scale in every accretion episode. 
Hence, the only variable in our model is the Eddington ratio.

Next, we use Millennium simulation which has simulation box five times larger
($L=500h^{-1}~\mathrm{Mpc}$) in order to compare final 
BH masses in merger trees of this simulation to the observed 
BH mass function.
We apply the same BH growth recipe to the haloes of Millennium simulation. 
The difference is, time the seed BH mass is the most common BH mass
estimated in the merger trees of Millennium-II simulation.

We use Millennium-II simulation to determine a typical BH mass for a 
specific host halo mass. We do it by binning masses of resulting BHs 
at each snapshot and halo mass interval and then choosing central value of the most occupied bin.
Hence, at every redshift we know the range of BH masses hosted by haloes 
with a specific mass.
We select the typical (most common mass) BH 
and use it as a BH seed in each newly formed halo of Millennium simulation at the matching redshift 
and halo mass obtained from Millennium-II, thus extending our sample to include highest mass haloes. 
Note that this is a conservative way to populate haloes since we are losing instances when BH mass is in reality more massive 
than the common one.

\subsection{Merger tree}

First, we use Millennium-II data base to select haloes with masses 
$>10^{10}\Msun$ at redshift $z=6.2$. For all selected
haloes we find their progenitor haloes, i.e. haloes that have merged at 
the previous snapshot to form selected haloes. Haloes have merged if they 
have the same descendant halo. We repeat this procedure up to redshift 
$z=23.79$ where first mergers are recorded.
Once we find all of the haloes 
that take part in formation of the selected haloes, we form merger 
trees for each of them.

Final haloes at redshift $z=6.2$ are called main haloes. Mergers of 
the main halo form the main branch of the merger tree. Other haloes that 
have merged with the main halo are side haloes and their previous 
mergers form side branches. Every merger tree contains main branch 
and side branches. We follow not only mergers of the main halo, but also 
take into account mergers in all side branches. 

In a single merger event the most massive progenitor halo 
is called primary halo and other progenitor haloes are satellite haloes.
In the Millennium-II simulation output 
is printed with time resolution $\Delta z\sim1$, so the common case is 
that large number of haloes have the same descendant halo (case with 
largest number of progenitors being 186). The question is which halo 
is the primary halo and which haloes are satellite haloes? To solve this 
we first find the most massive progenitor halo (primary halo) and then
using the comoving coordinates of the centre of mass, 
we calculate distances to other progenitor haloes (satellite haloes), sort 
them in the order from closest to furthest away and assume that they have 
merged in that order. We calculate halo mass ratio of the merging 
haloes and distinguish minor and major mergers.

We repeat the same procedure for haloes in Millennium simulation
that have masses $>10^{11}\Msun$ at redshift $z=6.2$.

\subsection{Parameters choice and initial BH masses}

Final BH mass depends on the initial BH mass, Eddington ratio, radiative 
efficiency and accretion time-scale (equation 2.). 
In our model, every major merger leads first 
to the formation of a new BH after which gas accretion phase is triggered.

Since we are only interested in the initial and the final BH mass before 
and after every accretion episode, we do not strictly insist on any particular 
accretion model. The accretion parameters in our model should be regarded as 
averaged over the accretion time-scale no matter what the actual BH growth model is. 
These averaged parameters could be the real accretion parameters in the classical
'thin disc' accretion model, or they could be an average of a sequence of short-lived 
intermittent phases of super-Eddington accretion (with Eddington ratios greater than the averaged one) 
in the 'slim disc' accretion model \citep{vs}. That is why we refer to Eddington ratio 
and radiative efficiency as effective Eddington ratio and effective radiative efficiency in our model.

For effective radiative efficiency 
we choose commonly accepted value $\epsilon=0.1$ (\citealt{elvis}; 
\citealt{yu}; \citealt{davis}). 
Every accretion episode is limited to 50 Myr, which is $\sim$ Salpeter's
time for accretion at the Eddington ratio $f_{\mathrm{Edd}}=1$.
The only free parameter in our model is the effective Eddington ratio.
We assign a fixed value of this parameter
to each accretion episode in one simulation run.
Accretion episode can be shorter than 50 Myr in two cases.
First, if a new major merger occurs before that time has passed, accretion is reset and new accretion 
episode begins. Secondly, if BH mass exceeds 0.08 per cent of the host halo mass, accretion will be stopped
and cannot be triggered again. 
This constraint comes from BH-dark matter halo mass scaling relation in the local Universe. 
Mass in baryons is approximately 16 percent 
of the dark matter mass. If a typical BH mass to 
bulge mass ratio is 0.5 per cent (\citealt{kh}, \citealt{novak}), then $8\times10^{-4}$  
of the halo mass is the gas that BH can accrete.

In each newly formed halo of Millennium-II simulation we 
place one BH with initial mass of $100\Msun$ and run
one simulation with fixed accretion parameters.
We use haloes of Millennium-II simulation to estimate what are the common masses
of BHs that populate haloes of a certain mass at each redshift.
Then we use those common BHs as seeds in newly formed haloes of Millennium simulation
and apply the same BH growth recipe with the same accretion parameters as we have 
used in Millennium-II simulation.

We use the observed BH mass function at $z\sim6$ as a constraint 
for our model. 
We run a set of semi-analytical simulations for different values of 
the effective Eddington ratio. In each run we assign the same values 
for the effective Eddington ratio to each accretion episode.
We make sure not to overproduce SMBH at $z=6.2$ (Millennium snapshot
closest to redshift $z=6$)
and once this condition is satisfied, we check if we have
$10^{9}\Msun$ SMBH at $z=7$ for the specific choice of the effective Eddington ratio.

\section{RESULTS}

We find that if the effective Eddington ratio is $f_{\mathrm{Edd}}=3.7$
both conditions of our model are satisfied: 
BH mass function is consistent with the observed BH mass function at $z\sim6$ \citep{massf}
and our merger tree produces $10^{9}\Msun$ SMBH at $z=7$.

Fig. 1 shows mass function of BHs that populate $>10^{11}\Msun$ haloes
of Millennium Simulation at $z=6.2$. BH masses are the result of one 
semi-analytic simulation in which we chose initial BH masses to be $100\Msun$, effective
radiative efficiency $\epsilon=0.1$,  effective Eddington ratio $f_{\mathrm{Edd}}=3.7$ and 
each accretion episode is limited to 50 Myr. 
In order to calculate BH mass function in our model 
we bin masses of all BHs in haloes of Millennium simulation at redshift z=6.2.
Points depicted by plus symbols in Fig. 1 represent numbers of BHs in each bin per Mpc$^{3}$.
Bins have width of 0.1 dex (in logarithmic scale).
We compare our BH mass function to the BH mass function at $z=6$ given by \cite{massf} (dashed and solid lines).
\cite{massf} modelled BH mass functions to produce luminosity
functions which are then fitted to the observed luminosity function
of Canada-France High-z Quasar Survey (CFHQS) and SDSS
quasars at $5.74<z<6.42$  \citep{lumf}.
Dashed lines in Fig. 1 represent their upper and lower limits,
while the solid line is their best fitting to the data.

\begin{figure}
\includegraphics[width=90mm]{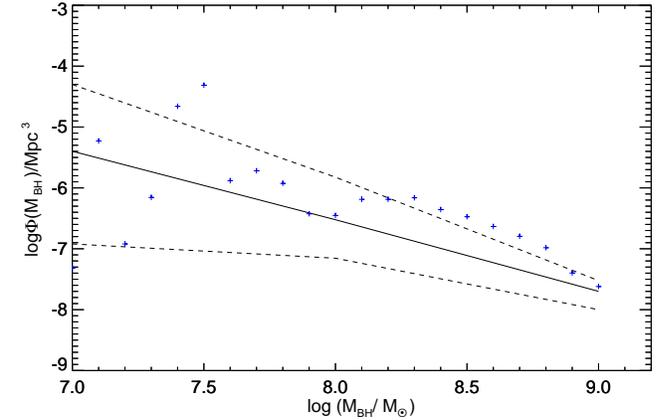}

\caption{Mass function of BHs in our model (blue plus symbols), compared to 
BH mass function given by Willott et al. (2010b). Dashed lines show upper and lower limit
from Willott et al. (2010b) while the solid line
shows their best-fitting.
}
  \label{sample-figure}
\end{figure}

\cite{shen} showed that accretion with similar Eddington ratios exists in high 
redshift quasars. They found that the maximum observed value of Eddington 
ratio for the observed quasars is $f_{\mathrm{Edd}}\sim3$. However, other authors 
suggest that even larger values of Eddington ratio are possible 
($f_{\mathrm{Edd}}=4.6$ \citep{du}, or even $f_{\mathrm{Edd}}=10$ \citep{collin}).

Assuming that a moderate super-Eddington 
accretion is possible for a prolonged period of time (more than one e-folding time), 
in the further analysis we focus on the main halo with the most massive SMBH 
at redshift $z=7$ and its progenitors.

\begin{figure}
\includegraphics[width=90mm]{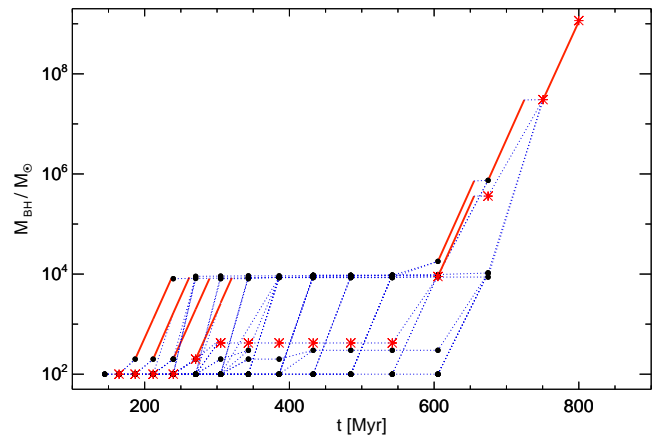}

\caption{Merger tree of a $10^{9}\Msun$ SMBH at redshift  
$z=7$ for $f_{\mathrm{Edd}}=3.7$. It follows the 
entire growth of a SMBH through all mergers and accretion episodes, as a 
function of the age of the Universe. Black circles represent BH masses in side haloes at the snapshot-times, 
while red asterisks represent BH masses in the main halo.
Dotted blue lines follow BH growth by minor mergers and solid red 
lines show major mergers which are always followed by the gas accretion.}
  \label{sample-figure}
\end{figure}

Fig. 2 shows the merger tree of a $10^{9}\Msun$ SMBH at 
redshift  $z=7$ and effective $f_{\mathrm{Edd}}=3.7$. It follows 
the entire growth of a SMBH through all mergers and accretion episodes, as a 
function of the age of the Universe. Black circles represent BH masses in side 
haloes (side branches) at the snapshot-times, while red asterisks represent BH masses in the main halo (main branch).
Dotted blue lines follow BH growth by minor mergers and solid red 
lines show growth through the gas accretion which occurs after every major merger. 
At the end, there is only one SMBH left. BH gains most of its mass in 
major mergers when accretion is triggered.

Accretion episodes at $\rm{t}<300$Myr occur in haloes of Millennium-II simulation,
while accretion episodes at $\rm{t}>600$Myr represent major mergers of haloes in Millennium simulation. 
The absence of accretion episodes between $\rm{t}=300$Myr and $\rm{t}=600$Myr is the consequence
of the fact that we do not know the exact merger history of low mass haloes of Millennium simulation.
We approximate BH masses in those haloes with typical BHs that populate haloes of
Millennium-II simulation at the matching redshift.

Each accretion episode lasts 50 Myr.
Salpeter's time for a BH accreting at the effective Eddington ratio $f_{\mathrm{Edd}}=3.7$ is
$\sim12$Myr (equation 3), which means that e-folding time in each accretion episode is $\sim4$.

Super-Eddington accretion 
for such a long period of time might seem unrealistic, but it could 
be more possible than the accretion at the Eddington limit for almost a billion 
years. 
To explain the formation of $>10^{9}\Msun$ SMBH with a former classical values of the accretion parameters 
($M_{\mathrm{BH,0}}=10^{5}~\mathrm{M_{\odot}}$, $\epsilon=0.1$ and  $f_{\mathrm{Edd}}=1$)
it can be calculated that continuous accretion for 500 Myr is needed (equation 2.) and e-folding time is $\sim11$.
If we start with those parameters and smallest BH seed of $100\Msun$, we calculate that e-folding
time is $\sim$19 and the accretion lasts for 840 Myr in case where
$f_{\mathrm{Edd}}=1$ and 280 Myr in case where $f_{\mathrm{Edd}}=3$.

In our model, growing SMBH at $z=7$ does not require e-folding
time larger than 4 for $100\Msun$ BH seeds. 
This is a consequence of accretion being restarted in major mergers.
Instead of one BH constantly accreting we have several BHs in shorter
accretion episodes, where each of them is triggered in a major merger. 
SMBH in the main halo
grows through two accretion episodes while the  
rest of the accretion occurs in side haloes (Fig. 2). 
Above $10^{4}\Msun$ two accretion episodes occur in a side halo. 
BH in that side halo grows parallel with BH in the main halo.
When the side and the main haloes merge side halo hosts more massive
BH than the main halo because it has greater number 
of major mergers in its history. 
This approach increases the impact 
of mergers, and reduces importance of accretion, which in turn alleviates
need for super-Eddington accretion with the large number of e-folding times.

\begin{figure}
\includegraphics[width=90mm]{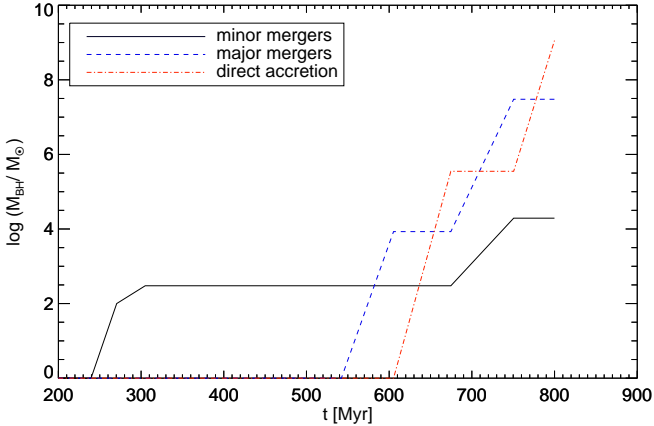}
 
\caption{
Growth of the SMBH in the main halo by
minor mergers (black solid line), major mergers (blue dashed line) 
and direct accretion (red dash-dotted line). Below $10^7\Msun$ 
major mergers and accretion are equally important
while above that mass SMBH growth is due to direct accretion.
}
\label{sample-figure}
\end{figure}

Fig. 3 represents accumulative contribution to the SMBH growth from
minor mergers (black solid line), major mergers (blue dashed line) and direct accretion (red dash-dotted line). 
It follows the growth of SMBH in the main halo. 
Major mergers have significant contribution  below $10^7\Msun$ 
because BHs in side haloes can grow to high masses before they merge 
with the BH in the main halo. Above $10^7\Msun$  SMBH growth is dominated
by direct accretion.

We also investigate if gravitational wave recoil could prevent the formation of SMBH. 
Fig. 4 shows the mass ratio of the merging BHs in the host 
haloes of $z=7$ merger tree. 
It shows the main and side haloes masses as a function of BH mass 
ratio. Our model might be sensitive to the choice 
of kick velocity. Most of the BH mergers with mass ratios close to one 
occur in low mass haloes. As low mass haloes have small gravitational 
potentials, final BH after the merger can easily be ejected. If BH survives
these mergers and settles in a larger halo, then the larger gravitational
potential will protect it from any following merger. 
Red diamonds represent haloes where the sum of two merging BHs 
is $>10^{7}\Msun$. Those mergers have the largest influence on the BH growth and since
they occur in large haloes and have low mass ratio, they are not affected by the
gravitational wave recoil.
Fig. 4 also shows 
that mass ratio of merging BHs decreases in larger mass haloes.

\begin{figure}
\includegraphics[width=90mm]{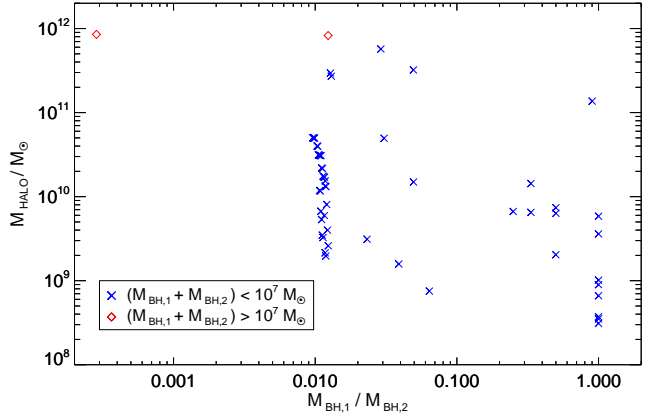}

\caption{Main and side haloes masses as a function of BHs mass ratio 
in both major and minor halo mergers. 
Red diamonds represent haloes where the sum of two merging BHs is $>10^{7}\Msun$.
Those mergers are the most important in BH growth and they are not affected by gravitational wave radiation.
Mergers of BHs with mass ratio $\sim1$ might suppress 
the growth of SMBH by gravitational wave recoil since they reside in low-mass haloes. BHs that form in larger
haloes are protected from ejection by the large gravitational potentials.}
\label{sample-figure}
\end{figure}

\begin{figure}
\includegraphics[width=90mm]{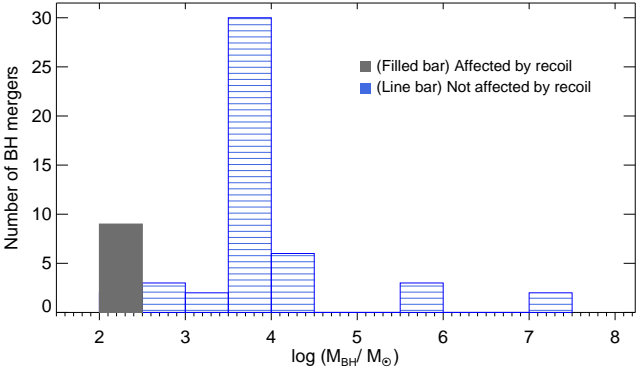}

\caption{Number of BH mergers as a function of BH mass. Grey filled histogram shows BHs whose
kick velocities are larger than the escape velocities from their host halo, so they are sensitive to 
the gravitational wave recoil. Blue histogram with horizontal lines shows BHs that are not affected by gravitational wave recoil.}
\label{sample-figure}
\end{figure}

To further investigate the influence of the gravitational wave recoil, 
we calculate and compare kick value with the value of the halo escape velocity.

Kick velocities are taken from Micic \etal (2011). 
The authors used parametrized fit of \cite{campa} to calculate
the kick velocity as a function of the merging BHs mass ratio,
BHs spin and the alignment to the orbital angular momentum.
Kick velocity is (their equations 7-10):
\begin{equation}
v_{\mathrm{kick}} = [(v_{m}+v_{\perp} cos{\xi})^2+(v_{\perp} \sin{\xi})^2+(v_{\parallel})^2]^{1/2},  
\end{equation}
where
\begin{equation}
v_{m} = A \frac{q^2 \left(1-q\right)}{\left(1+q\right)^5} \left[ 1+
  B \frac{q}{\left(1+q\right)^2}\right], 
\end{equation}
\begin{equation}
v_{\perp} = H \frac{q^2}{\left(1+q\right)^5} \left( \alpha_2^\parallel
  - q \alpha_1^\parallel\right), 
\end{equation}
and
\begin{equation}
v_{\parallel} = K \cos\left(\Theta-\Theta_0\right) \frac{q^2}{\left(1+q\right)^5}
\left( \alpha_2^\perp - q \alpha_1^\perp \right).
\end{equation}
\noindent The fitting constants are A$=1.2 \times 10^{4} \kms$, B$=-0.93$, H$=(7.3 \pm 0.3) \times 10^{3} \kms$
and K$\cos\left(\Theta-\Theta_0\right) =(6.0 \pm 0.1) \times 10^{4} \; $,
q is mass ratio of the merging BHs, $\alpha_i$ is the reduced spin parameter and 
the orientation of the merger is specified by angles $\Theta$ and $\xi$.
The authors show distribution of gravitational recoil for two models: when 
the spin parameters are chosen from a uniform distribution and when it is assumed
that the BH spins are aligned with the orbital angular momentum.
We accepted their model when BH spins are aligned with the orbital angular momentum (their fig. 2, red region)
and we used maximum of the kick velocity distribution (their fig. 2, red line) to 
assign values of kick velocities to the merging BHs in our model.

Escape velocity is calculated as described in \cite{vesc}. 
Circular velocity of a halo at the virial radius is \citep{barkana}:
\begin{equation}
 v_{\rm{c}}=24(\frac{M_{\rm{gal}}}{10^{8}h^{-1}\Msun})^{1/3}(\frac{\Omega_{\rm{m}}}{\Omega^{z}_{\rm{m}}}\frac{\Delta_{\rm{c}}}{18\pi^{2}})^{1/6}(\frac{1+z_{\rm{merge}}}{10})^{1/2} \rm{km s^{-1}}
\end{equation}
where $\Delta_{\rm{c}}=18\pi^{2}+82d-39d^{2}$, $d=\Omega^{z}_{\rm{m}}-1$ and $\Omega^{z}_{\rm{m}}=\Omega_{\rm{m}}(1+z)^{3}/(\Omega_{\rm{m}}(1+z)^{3}+\Omega_{\Lambda})$
at the merger redshift.  
The authors assumed that the dark matter haloes are described with NFW profile 
\citep{nfw} with a concentration parameter $c=4$ out to the virial
radius, as expected for a newly formed dark-matter halo \citep{wech}.
Under those assumptions the escape velocity from the halo's centre is $v_{\rm{esc}}\approx2.8v_{\rm{c}}$.

Fig. 5 shows the number of BH mergers as a function of the mass of the newly formed BH. 
If BH kick velocity is larger than the escape velocity from the host halo, 
newly formed BH could be ejected from the halo (grey filled histogram). Blue histograms with horizontal lines
represent BHs whose kick velocities are smaller than the escape velocities.
The figure shows that low-mass BHs reside in low mass haloes and have their kick velocities larger 
than the escape velocities from their host haloes, so they are sensitive 
to the gravitational wave recoil. On the other hand, we use simple model in 
which we assume that each newly formed halo in Millennium-II simulation
hosts a BH with the initial mass of $100\Msun$. Possibility that 
two merging BHs have exactly the same mass is very small. 
First BHs have 
initial mass function (IMF), with masses in wide range. IMF for Pop III BH remnants is not fully understood
due to uncertainty in primordial gas fragmentation during Pop III star formation. Pop III stars with masses 
in range $\sim25-140\Msun$ are believed to leave BH remnants with masses $M_{\rm{BH}}\sim10-50\Msun$ while 
more massive stars, $\gtrsim260\Msun$, leave BHs with masses $M_{\rm{BH}}\sim100-600\Msun$ (e.g., \citealt{hw}).
Such IMF would significantly reduce the mass ratio
of merging BHs. In turn this would lower the value of gravitational wave recoil.

\section{DISCUSSION AND CONCLUSIONS}

In this paper we test under which conditions light BH seeds (100 $\Msun$)
placed in haloes of Millennium-II simulation \citep{boy}
and Millennium simulation \citep{MS} merger trees 
can grow to SMBHs with masses $10^{9}\Msun$ that have been 
observed at $z\sim7$.

We make merger trees which track dark matter halo merger history from redshift $z=23.79$ 
to $z=6.2$. We assume that each halo hosts one BH and that two BHs merge 
right after their host haloes merge.
BH can grow in BH mergers and by gas accretion. We have distinguished between minor 
and major mergers. In the case of a 
minor merger, mass of the newly formed BH is a simple sum of the 
previous BH masses. Major merger leads 
to the formation of a new BH and it triggers the gas accretion on to that BH. 
Final BH mass depends on the initial BH mass, Eddington ratio, radiative 
efficiency and quasar lifetime (equation 2.). Effective
radiative efficiency is fixed at $\epsilon=0.1$ (\citealt{elvis}; 
\citealt{yu}; \citealt{davis}) and each accretion episode
is limited to 50 Myr, which is $\sim$ Salpeter's time for accretion at the Eddington limit.
Eddington ratio is a free parameter in our model, but it has fixed value
for each accretion episode in one simulation run.
If a new major merger occurs before the maximum allowed time for accretion  
has passed, accretion is reset and the new accretion episode begins.
Also, accretion can be stopped if BH mass exceeds 8 $\times$ 10$^{-4}$ of the host halo mass, so we
make sure that the gas reservoir never gets depleted.

We combine Millennium-II and Millennium merger trees in order to have both:
early halo formation history with low-mass haloes (to track BH growth history); 
and a large box giving us abundant highest mass haloes later (in which $10^9\Msun$ SMBH at $z\sim7$ 
can be produced). First we place $100\Msun$ BH seeds in haloes of Millennium-II simulation 
which has 125 times better mass resolution to make BH growth history.
Then we take most common BHs from Millennium-II simulation
as seeds for Millennium simulation to produce $10^9\Msun$ SMBH at $z\sim7$.

We run the set of semi-analytical simulations 
where we assign same initial masses of $100\Msun$ to all BH seeds and the same value 
for the effective Eddington ratio during the episodes of accretion.
We investigate what value of the effective Eddington ratio match two conditions:
BH mass function in our model cannot exceed BH mass function given by \cite{massf}
and our merger trees need to produce $10^9\Msun$ SMBH at $z=7$.

We find that remnants of Pop III stars can produce $10^{9}\Msun$ SMBH
at redshifts $z=7$ if at each accretion episode 
they are able to accrete at the effective Eddington ratio of
$f_{\mathrm{Edd}}=3.7$.
Recent observations have suggested that moderate super-Eddington accretion ($1<~f_{\mathrm{Edd}}<~10$)
might be possible (\citealt{shen}; \citealt{du}; \citealt{page}; \citealt{novak}). 
 
The question is then how long can accretion maintain rates above the Eddington limit?
Some authors argued that quasars lifetimes are much shorter then the Salpeter's time-scale 
(e.g, \citealt{rich}; \citealt{wyi}). 
In that case, even massive BH seeds do not have enough time to grow to $>10^{9}\Msun$ SMBH at redshift $z\gtrsim7$.
A single BH seed with mass $10^{6}\Msun$ requires nine e-folding times to produce $>10^{9}\Msun$
SMBH at $z=7$, while $100\Msun$ seed requires 19 e-folding times for the same SMBH. 
Because of this requirement some previous works rejected Pop III star remnants as possible 
candidates for SMBH seeds (e.g., \citealt{john1}). 
 
In our model, growing SMBH at $z=7$ does not require e-folding
time larger than $\sim4$ for $100\Msun$ BH seeds and effective $f_{\mathrm{Edd}}=3.7$.
Thus, we have managed to reduce the time that BH needs to spend accreting.
In our model, every major merger restarts the accretion, hence,
instead of one BH constantly accreting for a long time (large number of e-folding
times) we have several BHs in shorter (small number of e-folding times) accretion episodes.
This approach increases the importance of mergers on SMBH growth.

Our model requires prolonged super-Eddington accretion which could
produce strong feedback that stops the inflow of gas towards SMBH and disrupts the SMBH accretion long before 50 Myr.
This could occur if we would assume the classical 'thin disc' accretion.
However, we do not assume any accretion model in advance. The chosen values 
for the radiative efficiency and the Eddington ratio in our model should be
regarded as the effective values, averaged over 50 Myr. BH could also grow 
through a sequence of short-lived intermittent phases of super-Eddington accretion ($f_{\mathrm{Edd}}>3.7$)
without the feedback. Eddington ratio averaged over $50~\rm{Myr}$ would then be $f_{\mathrm{Edd}}=3.7$ and effectively 
this model would produce the same SMBH as in the 'thin disc' model. The absence of feedback in the 
'slim disc' model comes from $f_{\mathrm{Edd}}=\epsilon \frac{\dot{M}}{\dot{M}_{\mathrm{Edd}}}$. 
BH mass growth rate could stay the same as radiative 
efficiency decreases and $f_{\mathrm{Edd}}$ increases. In principle, given a sufficiently low efficiency, a
super-Eddington BH may be emitting at sub-Eddington luminosity, thus eliminating BH feedback completely \citep{vs}. 

{\bf In light of the recent observations of super-Eddington accretion mentioned above, we show that 
with a moderate super-Eddington accretion $\boldsymbol{(f_{\mathrm{Edd}}=3.7)}$ averaged over 50 Myr, even low mass 
seeds $\boldsymbol{(100\Msun)}$ could be progenitors of high redshift SMBHs $\boldsymbol{(z \sim 7)}$.}

Note that our model is conservative for three reasons:

(1) All BH seeds in our model have the same mass of $100\Msun$. It is possible that Pop III stars 
left BH seeds with masses up to $300\Msun$ \citep{volker}, or even 
$1000\Msun$ \citep{hirano}. Larger mass of BH seeds would significantly reduce need for accretion
which would, in turn, effectively lower the value of the effective Eddington ratio.

(2) We combine two simulations which is also a conservative approximation. We do not know
the exact merger history of low mass haloes in Millennium simulation, but instead we 
take it from Millennium-II simulation. We always chose the most common BHs from Millennium-II
simulation to be the seeds for newly formed haloes of Millennium Simulation. 
If we would have chosen just a few BHs with mass greater than the most common mass, 
that would strongly affect BH growth history. This in turn would reduce the 
Eddington ratio necessary for SMBH growth.

(3) In our model BH growth is limited by the amount of gas that BH can accrete ($8\times10^{-4}$ 
of the host halo mass). This constraint comes from the BH mass scaling relations in the local Universe.
However, the same scaling relation may not hold in the early Universe.
Recently, \cite{barn} measured $\dot{M}_{\textrm{bh}}/\dot{M}_{\textrm{bulge}}\backsimeq0.2$
for quasar ULAS J1120+0641 at $z=7.1$. They found that the BH was growing
much faster than the bulge relative to the mass ratio measured in the local Universe.
If BHs at high redshift had more gas available for the accretion,
BH growth would be much easier which would also reduce the value of the effective Eddington ratio.

Some authors have speculated that BH mergers have a limited role in SMBH growth due to gravitational wave recoil
(e.g., \citealt{merritt}; \citealt{volonteri2007}). We calculate BH kick velocities and 
compare it to the escape velocities of their host haloes. We find that BHs with low masses 
reside in low mass haloes and have their kick velocities larger 
than the escape velocities from their host haloes, so they are sensitive 
to the gravitational wave recoil. This is a consequence of a simple model where all BH seeds
have the same mass which is highly unlikely. Using a proper initial BH mass function 
with a wide range of possible masses would lower the value of gravitational wave recoil.

We note that, in order to get more accurate results, our model should be applied to
the merger trees of a simulation that would have both resolution of Millennium-II simulation,
and the box size as in Millennium simulation.

\section{ACKNOWLEDGEMENTS}
        
This work was supported by the Ministry of Education, Science and Technological Development of the
Republic of Serbia through project no. 176021, 'Visible and Invisible Matter in Nearby Galaxies:
Theory and Observations'.

\label{lastpage}

\end{document}